\documentclass[superscriptaddress,preprintnumbers,amssymb,nofootinbib]{revtex4-2}

\usepackage{tikz-feynman}
\usepackage{subfig}
\usepackage{bm}
\usepackage{esvect}
\usepackage{verbatim}
\usepackage{multirow}
\usepackage[left]{lineno}
\usepackage{blindtext}
\pdfoutput=1
\usepackage[T1]{fontenc}
\usepackage{graphicx}
\usepackage{epsfig}
\usepackage{amsmath}
\usepackage{amsfonts}
\usepackage{amssymb}
\usepackage{braket}
\usepackage{cancel}
\usepackage{pstricks}
\usepackage{color}
\usepackage{feynmf}
\usepackage[normalem]{ulem}
\usepackage{epstopdf}
\usepackage{soul}
\usepackage{cancel}

\definecolor{ao(english)}{rgb}{0.0, 0.5, 0.0}

\usepackage{braket}
\usepackage{cancel}
\usepackage{setspace}
\usepackage{pstricks}
\usepackage{xcolor}
\usepackage{float}
\usepackage{mathtools}
\usepackage{multirow}
\usepackage{booktabs}
\usepackage{enumitem}
\providecommand{\tabularnewline}{\\}

\def\gsim{\lower0.5ex\hbox{$\:\buildrel >\over\sim\:$}}
\def\lsim{\lower0.5ex\hbox{$\:\buildrel <\over\sim\:$}}

\def\tp{{\cal O}_{\tt CP}}

\def\be{\begin{equation}}
\def\ee{\end{equation}}
\def\bea{\begin{eqnarray}}
\def\eea{\end{eqnarray}}

\def\op{Q}

\begin{document}

\title{Theoretical underpinnings of CP-Violation at the High-energy Frontier}

\author{Shaouly Bar-Shalom}
\email{shaouly@physics.technion.ac.il}
\affiliation{Physics Department, Technion--Institute of Technology, Haifa 3200003, Israel}
\author{Amarjit Soni}
\email{adlersoni@gmail.com}
\affiliation{Physics Department, Brookhaven National Laboratory, Upton, NY 11973, USA}
\author{Jose Wudka}
\email{jose.wudka@ucr.edu}
\affiliation{Physics Department, University of California, Riverside, CA 92521, USA}

\date{\today}

\begin{abstract}
We present a general analysis for the discovery potential of CP-violation (CPV) searches in scattering processes at TeV-scale colliders in an effective field theory framework, using the SMEFT basis for higher dimensional operators. 
In particular, we systematically examine the CP-violating sector of the SMEFT framework in some well motivated limiting cases, based on flavour symmetries of the underlying heavy theory. We show that, under naturality arguments of the underlying new physics (NP) and in the absence of (or suppressed) flavour-changing interactions, there is only a single operator, $\op_{t\phi} = \phi^\dagger \phi  \left(\bar q_3 t \right) \tilde{\phi} $ which alters the top-Yukawa coupling, that can generate a non-vanishing CP-violating effect from tree-level SM$\times$NP interference terms. We find, however, that CPV from $\op_{t\phi} = \phi^\dagger \phi  \left(\bar q_3 t \right) \tilde{\phi} $ is expected to be at best of $O(1\%)$ and, therefore, very challenging if at all measurable at the LHC or other future high-energy colliders. 
We then conclude that a potentially measurable CP-violating effect of $O(10\%)$ can arise in high-energy scattering processes {\it only} if flavour-changing interactions are present in the underlying NP; in this case a sizable CPV can be generated at the tree-level by pure NP$\times$NP effects and not from SM$\times$NP interference. We provide several examples of CPV at the LHC and at a future $e^+e^-$ collider to support these statements. 
\end{abstract}

\maketitle
\flushbottom

\newpage 

\section{Introduction \label{sec:intro}}

One of the most important components of particle physics is CP-violation (CPV), since it is a key ingredient of the evolution of the universe and of the cosmological model \cite{Sakharov,Kuzmin:1985mm,Branco:2011zb}. 
It has been observed 
in 
$K, D$ and $B$ meson mixing and decays, where all observed CP asymmetries  in those
systems are found to be consistent with the Kobayashi-Maskawa mechanism for CPV in the Standard Model (SM) \cite{PDG}, which, however, 
seems  
insufficient for explaining the observed baryon asymmetry of the universe (BAU), see e.g., \cite{Rubakov:1996vz,Bernreuther:2002uj,Canetti:2012zc}.
An extremely important consequence of these observations of CP violation is that 
CP is therefore not a symmetry of nature. It is,therefore, natural to expect new Beyond the SM (BSM)
CP-odd phase(s).

It is for that reason that searches for new BSM CPV sources may be the pathway to a deeper 
understanding of particle physics and of the evolution of the observed universe. Given that no new physics (NP) has been detected yet up to the TeV scale, an important testing ground for BSM CPV effects is high-energy (multi-TeV scale) colliders, since CPV effects from an underlying heavy NP is suppressed by inverse powers of its scale. In particular, the NP that underlies the SM can be parameterized in general by higher dimensional, gauge-invariant effective operators, $\op_i^{(n)}$, in the so-called SM Effective Field Theory (SMEFT) framework, where the effective operators are constructed using the SM fields and their coefficients are suppressed by inverse powers of the NP scale $\Lambda$
\cite{EFT1,EFT2,EFT3,EFT4,EFT5}:
\begin{eqnarray}
{\cal L} = {\cal L}_{SM} + \sum_{n=5}^\infty
\frac{1}{\Lambda^{n-4}} \sum_i \alpha_i \op_i^{(n)} \label{eq:EFT1}~,
\end{eqnarray}
where $n$ is the mass dimension of $\op_i^{(n)}$ and we assume decoupling and weakly-coupled heavy NP, so that
$n$ equals the canonical dimension. 
One can further divide the higher-dimension effective operators to those that can be potentially generated at tree-level (PTG) 
and/or are loop generated (LG) in the underlying heavy theory \cite{jose_PTG_LG}.   
The dominating NP effects are then expected to be generated by contributing operators with the 
lowest dimension (smallest $n$) that are PTG in the underlying heavy theory. 
Furthermore, the (Wilson) coefficients $\alpha_i$ depend on the details of the underlying heavy theory and, therefore, 
they parameterize all possible weakly-interacting and decoupling types of heavy physics. In particular, 
it is expected that  $\alpha_i =O(1)$ for the PTG operators and  $\alpha_i \sim 1/(4\pi)^2$ if the operators are LG; the LG operators are thus a-priory suppressed by a loop factor and, therefore, their effects at lower energies (i.e., $E < \Lambda$) are expected to be subleading.  

In this paper we will characterize and identify the possible manifestations of new CPV sources from generic decoupling and weakly-coupled underlying heavy NP, using the SMEFT framework and exploiting the properties of the SM and SMEFT framework as well as the possible flavor structures of the underlying heavy physics. 
We will show that high-energy scattering processes involving the top-quark are the best and possibly {\it only} candidates for hunting BSM sources of CPV. Indeed, the top-quark, because of its relatively large mass, is expected to be more sensitive to TeV-scale NP 
and to 
CPV from BSM sources \cite{ourreview}; besides, CPV effects in top-quark systems from the SM CKM CP-odd phase 
are expected to be un-observably small \cite{ourreview}. 

In particular, we show that if flavor violation is suppressed in the underlying heavy theory, then only a single operator that involves the top-quark and Higgs fields can potentially yield non-negligible CPV which is, however, still below the expected sensitivity of the LHC and/or future high-energy colliders. On the other hand, if the underlying heavy theory contains flavor violating $t \to u$ and/or $t \to c$ transitions, then sizable tree-level CPV of $O(10\%)$ can be generated, that is within the reach of e.g., the LHC.   

\section{Patters of CP-violation in the SMEFT} 

We divide the CPV SMEFT sector into two groups: CPV operators that can generate $O(1/\Lambda^2)$  interference effects with the SM,~\footnote{We will ignore the single dimension 5 operator in SMEFT since it is lepton-number violating and its scale is accordingly expected to be very high; we also ignore operators of dimension $ \ge7$ since they are suppressed by higher powers of the NP scale $ \Lambda $.} and those that cannot interfere with the SM to this order when CKM suppression factors are ignored. The first group can be further subdivided depending on whether the interference with the SM is proportional to the fermion masses or is not, while the second group consists of flavor-changing operators.\footnote{Other classifications of the CP-odd sector in the SMEFT framework, based on similar arguments, can be found in \cite{SMEFTsim2,2110.02993,2112.03889}.}   

\subsection{CPV from interference with the SM \label{secIIA}}

In order to understand the patterns of CPV through interference with the SM, it is useful to consider two limiting cases of the SM that are well motivated phenomenologically and experimentally: 
\begin{description}
\item[\boldmath{${\rm SM}_0$}] all masses in the SM are set to zero, so that the fermion sector is segregated into a number of non-interacting sub-sectors, each labeled by its flavor and chirality. This theory has a ${\cal G}_0 = U(3)^5$ global flavor symmetry (without right-handed neutrinos) and it is CP-invariant (except for a possible strong CP $\theta$-term \cite{teta_term} that we ignore). Note that this limit motivated the so-called Minimal Flavor Violation (MFV) ansatz for model building of NP beyond the SM, where  the Yukawa couplings are promoted to spurion fields \cite{MFV1,MFV2}. 
\item[\boldmath{${\rm SM}_t$}] The ${\rm SM}_0$ with a massive top-quark. That is:
\begin{eqnarray}
{\cal L}_{{\rm SM}_t} = {\cal L}_{{\rm SM}_0} + \left( y_t {\bar q}_3 t \tilde\phi + {\rm H.c.} \right) ~, \label{LSMt}
\end{eqnarray}
where $q_3$ denotes the third-generation left-handed quark isodoublet, $t$ the right-handed top-quark isosinglet and $\phi$ is the SM Higgs field.
The ${\rm SM}_t$ has a reduced symmetry ${\cal G}_t = U(3)^4 \times U(2) \times U(1) \subset {\cal G}_0 $ and it can give rise to ``new'' processes compared to those described by the ${\rm SM}_0$, with amplitudes proportional to a power of $ y_t$.
\end{description}

The higher dimensional effective operators of the SMEFT can likewise be segregated into the sectors that posses the ${\cal G}_0$ and ${\cal G}_t$ global symmetries of the ${\rm SM}_0$ and  ${\rm SM}_t$, respectively. We denote by ${\rm SMEFT}_0$ the terms in SMEFT that are invariant under ${\cal G}_0$ and by ${\rm  SMEFT}_t$ those that are invariant under ${\cal G}_t$ but not contained in ${\rm SMEFT}_0$. 
Note that, in contrast with SM$_0$, SMEFT$_0$ contains CPV interactions (see below).
The corresponding ${\cal G}_0$ and ${\cal G}_t$-symmetric effective Lagrangians are then:
\begin{eqnarray}
{\cal L}_{{\cal G}_0} &=& {\cal L}_{{\rm SM}_0} + {\cal L}_{{\rm SMEFT}_0} \label{eq:G0} ~, \nonumber \\
{\cal L}_{{\cal G}_t} &=& {\cal L}_{{\rm SM}_t} + {\cal L}_{{\rm SMEFT}_0} + {\cal L}_{{\rm SMEFT}_t}  \label{eq:Gt} ~.
\end{eqnarray}

It then follows that only operators in ${\rm SMEFT}_0$ will interfere with the ${\rm SM}_0$. This implies that CPV interference effects between the SMEFT and the SM that are generated by operators {\em not} in ${\rm SMEFT}_0$ will be suppressed by at least one power of a fermion mass; of these the dominant ones will be those for which this suppression is $ \propto m_t $, and these are generated by  ${\rm SMEFT}_t$. 

In Table \ref{tab:SM0} we list the complete set of potentially CPV dimension six effective operators of the ${\rm SMEFT}_0$ and ${\rm SMEFT}_t$ sectors. 
Note that the CPV operators of the ${\rm SMEFT}_0$ sector do not involve fermion fields 
and they are all LG in the underlying heavy theory.  In the ${\rm SMEFT}_t$ sector, only the operator $\op_{t\phi}$ is PTG. Moreover, the LG 
operators of the ${\rm SMEFT}_0$ sector (as well as the operator $\op_{tG}$ in the ${\rm SMEFT}_t$ sector) are tightly constrained by the lepton and neutron electric dipole moments \cite{2109.15085}.
It thus follows that the leading CPV effect from interference with the SM is generated by $\op_{t\phi}$, which is the only PTG operator in the ${\rm SMEFT}_t$ sector; all other operators that interfere with the SM will yield CPV which is either loop suppressed (from LG operators) or suppressed by powers of light fermion masses.

\begin{table*}[htb]
\caption{\label{tab:SM0}
The dimension six CPV effective operators of the ${\rm SMEFT}_0$ and ${\rm SMEFT}_t$ sectors, where $q_3$ is the left-handed SU(2) doublet of the 3rd generation
quarks and $t$ is the right-handed 
top-quark singlet.}
\renewcommand{\arraystretch}{1.5}
\begin{tabular}[t]{c|c}
${\cal L}_{{\rm SMEFT}_0}^{\rm CPV}$ & ${\cal L}_{{\rm SMEFT}_t}^{\rm CPV}$ \\

\hline
$\op_{\tilde{G}} = f^{ABC} \tilde{G}_\mu^{A\nu} G_\nu^{B\rho} G_\rho^{C\mu}$ & $\op_{tG} = \left(\bar q_3 \sigma^{\mu \nu} T^A t \right) \tilde{\phi} G_{\mu \nu}^{A} $ \\
$\op_{\tilde{W}} =f^{IJK} \tilde{G}_\mu^{I\nu} W_\nu^{J\rho} W_\rho^{K\mu}$ & $\op_{tW} = \left(\bar q_3 \sigma^{\mu \nu} t \right) \tau^I \tilde{\phi} W_{\mu \nu}^{I} $ \\
$\op_{\phi\tilde{G}} = \phi^\dagger \phi  \tilde{G}_{\mu \nu}^{A} G^{A \mu \nu}$ & $\op_{tB} = \left(\bar q_3 \sigma^{\mu \nu} t \right)  \tilde{\phi} B_{\mu \nu} $ \\
$\op_{\phi\tilde{W}} = \phi^\dagger \phi  \tilde{W}_{\mu \nu}^{I} G^{I \mu \nu}$ & $\op_{t\phi} = \phi^\dagger \phi  \left(\bar q_3 t \right) \tilde{\phi} $ \\
$\op_{\phi\tilde{B}} = \phi^\dagger \phi  \tilde{B}_{\mu \nu} B^{\mu \nu}$ & \\
$\op_{\phi\tilde{W}B} = \phi^\dagger \tau^I \phi  \tilde{W}_{\mu \nu}^{I} B^{\mu \nu}$ & 
\end{tabular}
\end{table*}

\subsection{CPV from NP$\times$NP terms}

All non-Hermitian operators will have in general complex Wilson coefficients and can, therefore, generate CPV effects. As explained above, if the operators do not belong to the SMEFT$_{0}$ and SMEFT$_{t}$ sectors, then CPV from SM$\times$NP interference effects are negligible, but CPV from NP$\times$NP effects may still be significant.

In particular, it is worth listing here the non-hermitian operators that can in principle carry a CP-odd phase without flavor-violation in the underlying physics. These include the flavor-diagonal operators in ${\cal L}_{{\rm SMEFT}_t}^{\rm CPV}$ (see Table \ref{tab:SM0}) extended to all fermion species
as well as the following PTG flavor-diagonal operators:
\begin{eqnarray}
\op_{\phi u d} &=& i \left( \tilde\phi^\dagger D_{\mu} \phi \right) \left( \bar u \gamma^{\mu} d \right) ~, \nonumber \\
\op_{\ell e d q} &=& \left( \bar\ell^j e \right) \left( \bar d q_j \right) ~, \nonumber \\
\op_{\ell e q u}^{(1)} &=& \left( \bar\ell^j e \right) \epsilon_{jk} \left( \bar q^k u \right) ~, \nonumber \\
\op_{\ell e q u}^{(3)} &=& \left( \bar\ell^j \sigma_{\mu \nu} e \right) \epsilon_{jk} \left( \bar q^k \sigma^{\mu \nu} u \right) ~, \nonumber \\
\op_{q u q d}^{(1)} &=& \left( \bar q^j u \right) \epsilon_{jk} \left( \bar q^k d \right) ~, \nonumber \\
\op_{q u q d}^{(8)} &=& \left( \bar q^j T^a u \right) \epsilon_{jk} \left( \bar q^k T^a d \right) ~, \label{nonH}
\end{eqnarray}
where $j,k$ are SU(2) indices and $\ell$, $e$ denote the lepton fields in same generation and, similarly, $q,u,d$ are quark fields of a single generation. These operators do not conserve the global ${\cal G}_0$ and ${\cal G}_t$ symmetries mentioned above 
so that their interference effects  with the SM are suppressed by powers of $m/\Lambda$, where $m$ is a {\em light} fermion mass and, therefore, they are unobservable
(this is the case for both CPV and CP-conserving (CPC) reactions). The leading CPV effects in this case will potentially arise from interference between different NP contributions, which we denote as NP$^2$ effects that are $O(1/\Lambda^4)$ (this is also the case if the underlying heavy theory induces flavor-violating SMEFT operators), 
which limits the sensitivity of the LHC to such effects (see below).

Finally, if flavor is violated in the underlying heavy theory, then all SMEFT operators containing flavor-violating combinations of fermion fields can violate CP. These operators cannot interfere at tree-level with the SM (when the small off-diagonal CKM effects are ignored), so that the leading CPV effects in this case will again be generated from NP$^2$ terms, e.g., CPV effects from 4-fermion operators with generic flavor indices that we discuss in some detail below.

\section{CP-violation in high-energy scattering processes} 

The differential cross-section for any given scattering process may be divided into its CPV and CPC parts:
\begin{eqnarray}
    d \sigma \equiv d \sigma_{\rm CPC} + d \sigma_{\rm CPV}   ~.
     \label{sig}
\end{eqnarray}
In terms of these, the generic form of an observable sensitive to the corresponding CPV effects will be necessarily proportional to the ratio between the CPV and CPC terms:
\begin{eqnarray}
{\cal A}_{\rm CP} \propto \frac{d\sigma_{\rm CPV}}{d\sigma_{\rm CPC}} ~, \label{ACP_gen}
\end{eqnarray}      
so that it scales with the CPV contribution to the cross-section, $d\sigma_{\rm CPV}$, while being suppressed by the CPC terms of the corresponding 
cross-section, $d\sigma_{\rm CPC}$. 

Now, in order for a processes to violate CP, there must be  
at least two contributing amplitudes with different phases:
\begin{eqnarray}
    {\cal M}_{i \to f} &=& M_1 e^{i \left(\phi_1 + \delta_1 \right) } + M_2 e^{i\left( \phi_2 + \delta_2 \right) }  \label{M} ~,
\end{eqnarray}
where we have factored out the CP-odd and CP-even phases, $\phi_{1,2}$ and $\delta_{1,2}$, respectively; the latter arise from final state interactions (FSI) at higher loop orders (when no resonances are involved, which is the working assumption within the SMEFT framework). If the process under consideration is not self-conjugate, then the amplitude for the charge-conjugate (CC) channel (${\cal M}_{\bar{i} \to \bar{f}}$) is obtained from \eqref{M} by changing the sign of the CP-odd phases $\phi_i \to -\phi_i$ and replacing $M_i \to M_i^\star$. The CC reaction is useful in constructing CPV observables in the presence of CP-even phases (see Sect. \ref{subsec:CPV1}).

We are interested in studying the leading CPV effects, so we will  consider below only the cases where both $M_1$ and $M_2$ are generated at  tree-level,\footnote{If $M_i$ is generated at 1-loop it will be suppressed by an additional loop  factor $\sim 1/(16 \pi^2)$, with respect to the tree-level case.} in which case the CP-even phases vanish: $\delta_i =0$.

\subsection{Patterns of CP-Violation \label{CPVtypes}}

When considering only tree-level effects (i.e., no CP-even phases from FSI) and ignoring light quark masses, the general form of the amplitude, in the presence of the dim.6 SMEFT operators, takes the form
\begin{eqnarray}
{\cal M}_{i \to f} &= M_{\tt SM} +  \sum_k \frac{|\alpha_k|}{\Lambda^2} M_k e^{i \phi_k   } + \cdots \label{amplitude}
\end{eqnarray}
where $\alpha_k$ are the Wilson coefficient defined in Eq.~\ref{eq:EFT1} and the CPV phases ($\phi_k$) in $\alpha_k$ are factored out and explicitly displayed. The ellipsis denote contributions from higher dimensional operators.

Using the generic form of the NP amplitude in Eq.~\ref{amplitude}, we can now categorize {\it all possible} tree-level CPV manifestations that can be generated by the SMEFT in high-energy scattering processes, and estimate their magnitude. Recall that if the underlying heavy NP is natural, then we expect $\alpha_k \sim 1$ for PTG operators and $\alpha_k \sim 1/(16 \pi^2)$ for LG operators \cite{jose_PTG_LG}. Therefore, the leading CPV effects are expected to arise from PTG operators, and they can be of 3 types:
\begin{itemize}
\item{\bf TLCPV-I:} If there is SM-SMEFT interference. 
In this case, we argued above that there is only one PTG operator that can contribute, $ \op_{t\phi}$, so that (the ellipsis denote contributions $ \propto 1/\Lambda^4$)
\begin{align}
    d \sigma_{\rm CPC} &\propto \left| M_{\tt SM} \right|^2 +  \frac{|\alpha_{t\phi}|}{\Lambda^2} {\rm Re} \left(M_{\tt SM} M_{t\phi}^\dagger \right) \cos\phi_{t\phi}+ \cdots ~, \cr
    d \sigma_{\rm CPV} &\propto  \frac{|\alpha_{t\phi}|}{\Lambda^2} {\rm Im} \left(M_{\tt SM} M_{t\phi}^\dagger \right) \sin\phi_{t\phi}+ \cdots
\end{align}
\item{\bf TLCPV-II:} If there is no SM-SMEFT interference (for dimension 6 operators and higher). In this case, the  $1/\Lambda^4$ terms can be consistently retained (the ellipsis now denote terms $ \propto 1/\Lambda^6$):
\begin{align}
    d \sigma_{\rm CPC} &\propto \left| M_{\tt SM} \right|^2 + \sum_k \frac{|\alpha_k|^2}{\Lambda^4} |M_k|^2  + \sum_{k<l} \frac{|\alpha_k\alpha_l|}{\Lambda^4}  {\rm Re} \left(M_k M_l^\dagger \right) \cos\Delta\phi_{kl} + \cdots \,; \quad \Delta\phi_{kl} = \phi_k - \phi_l ~, \cr
    d \sigma_{\rm CPV} &\propto  \sum_{k<l} \frac{|\alpha_k\alpha_l|}{\Lambda^4}  {\rm Im} \left(M_k M_l^\dagger \right) \sin\Delta\phi_{kl} + \cdots
\end{align}
\item{\bf TLCPV-III:} If there is no SM contribution; so the relevant expressions are the previous ones with $M_{SM}=0$.
\end{itemize}

The above expressions, in all cases, take the generic form ({\it cf.} Eq. \ref{M})

\begin{eqnarray}
    d \sigma_{\rm CPC} &=& c_1 | M_1|^2 + c_2 | M_2|^2  +  c_3 {\rm Re} \left(M_1 M_2^\dagger \right) \cos\Delta\phi ~, \label{sigCPC} \\ 
     d \sigma_{\rm CPV} &=& c_{4} {\rm Im} \left(M_1 M_2^\dagger \right) \sin\Delta\phi  ~, \label{sigCPV1}
\end{eqnarray}
so that $\Delta\phi$ denotes a CP-odd phase and the coefficients $c_i$ have the following $\Lambda$ dependence: 
\begin{center}
\renewcommand{\arraystretch}{1.5}
\begin{tabular}[t]{c|c|c|c|c}
CPV type  &$~~c_1~~$ &  $~~c_2~~$ & $~~c_3~~$ & $~~c_4~~$ \\
\hline
TLCPV-I & $1$ & $  \frac{v^4}{\Lambda^4}$ & $\frac{v^2}{\Lambda^2}  $ & $\frac{v^2}{\Lambda^2} $ \\
TLCPV-II & $1$ & $  \frac{v_E^4}{\Lambda^4}$,  & $\frac{v^4_E}{\Lambda^4} $ & $\frac{v^4_E}{\Lambda^4} $ \\
TLCPV-III & $\frac{v^4_E}{\Lambda^4} $ & $\frac{v^4_E}{\Lambda^4} $ & $\frac{v^4_E}{\Lambda^4} $ & $\frac{v^4_E}{\Lambda^4} $
\end{tabular} 
\end{center}
where $v_E = v$ or $v_E=E$, depending on whether the SMEFT contributions give rise to an energy growing cross-section.

Also, at tree-level (i.e., in the absence of FSI), the imaginary part of the interference term, ${\rm Im} \left(M_1 M_2^\dagger \right) $, is necessarily proportional to a Levi-Civita tensor ($\epsilon_{\alpha \beta \gamma \delta}$) involving four independent momenta, so that the CPV term of the cross-section is in general: 
\begin{eqnarray}
   d\sigma_{\rm CPV} \propto c_{4}  \cdot \epsilon \left( u_{1},u_{2},u_{3},u_{4} \right) \sin\Delta\phi ~,
   \label{sigCPV2}
\end{eqnarray}
where the $u_i$ represent momenta or polarization/spin 4-vectors associated with the particles in the scattering process of interest and $\epsilon \left( u_1,u_2,u_3,u_4 \right) = \epsilon_{\alpha \beta \gamma \delta} u_1^\alpha u_2^\beta u_3^\gamma u_4^\delta$. Note that $\epsilon \left( u_1,u_2,u_3,u_4 \right)$ can always be expressed as a triple product $\vec u_i \cdot (\vec u_j \times \vec u_k)$ in some rest frame. Furthermore, to obtain a non-vanishing  $\epsilon \left( u_1,u_2,u_3,u_4 \right)$ requires at least 4 independent vectors, so that either there are at least 3 particles in the final state, i.e., $a + b \to 1 + 2+3$, or one must trace the spin/polarization information in a $2 \to 2$ scattering process.

A useful measure of the statistical significance, $N_{SD}$, with which an asymmetry ${\cal A}$ can be detected in a given process with cross-section $\sigma$ (ignoring systematic uncertainties) is provided by: 
\begin{eqnarray}
N_{SD}({\cal A}) \sim {\cal A}  \sqrt{\sigma \cdot {\cal L}} ~,
\label{NSD}
\end{eqnarray}    
where ${\cal L}$ is the available integrated luminosity. Referring to Eqs. \ref{M}, \ref{sigCPC}, \ref{sigCPV1} and \ref{NSD}, we summarize in Table ~\ref{tab:CPV0} the characteristics and $\Lambda$ dependence of the above three tree-level CPV scenarios, also addressing their potential observability $N_{SD}$. Evidently, 
scattering process with no SM contribution, where the NP yields a growing cross-section (TLCPV-III scenario) , i.e., $\sigma \propto E^4/\Lambda^4$, are expected to be the most sensitive to BSM CPV effects. An example of such a process is given in Sect. \ref{FCNP_CPV}. 
\begin{table*}[htb]
\caption{The $\Lambda$ dependence of the CP-asymmetry ${\cal A}_{CP}$ and its  potential observability $N_{SD}({\cal A_{CP}})$, as well as a description of the leading contributions to the CPV and CPC parts of the cross-section $\sigma$, for the three tree-level CPV scenarios TLCPV-I,II,III, where $v_E = v$ or $v_E =E$, depending on whether or not the SMEFT contributions give rise to an energy growing cross-section. See also text. \label{tab:CPV0}}
\renewcommand{\arraystretch}{1.5}
\begin{tabular}[t]{c|c|c|c|c}
 &   CPV source in $\sigma$& Leading CPC term in $\sigma$ & ${\cal A}_{CP}$ &  $N_{SD}({\cal A_{CP}})$  \\
\hline
TLCPV-I & ${\rm Im} \left(M_{SM} M_{NP}^{\dagger} \right) \propto \frac{v^2}{\Lambda^2} $ & $\left| M_{\tt SM} \right|^2$ & $\frac{v^2}{\Lambda^2}$ &  $\frac{v^2}{\Lambda^2}$ \\
TLCPV-II & ${\rm Im} \left(M_{NP'} M_{NP}^{\dagger} \right) \propto \frac{v_E^4}{\Lambda^4} $ & $\left| M_{\tt SM} \right|^2$ & $\frac{v^4_E}{\Lambda^4}$ &  $\frac{v^4_E}{\Lambda^4}$ \\
TLCPV-III & ${\rm Im} \left(M_{NP}'M_{NP}^{\dagger} \right) \propto \frac{v_E^4}{\Lambda^4} $ & $\left| M_{\tt NP} \right|^2 \propto \frac{v^4_E}{\Lambda^4}$ & $1$ &  $\frac{v^2_E}{\Lambda^2}$
\end{tabular}
\end{table*}

\subsection{Constructing observables for tree-level CP-Violation}\label{subsec:CPV1}

We now consider $2\to3$ scattering processes of the form $a b \to 1+ 2 +3$ (and the CC reaction $\bar{a} \bar{b} \to \bar{1} + \bar{2} + \bar{3}$) that are sensitive to CPV effects, and for which CPV observables  can be constructed without the need to analyze the spin/helicity of the final particles. We note that, in general, CPC contributions to the cross sections that depend on the CP-even phases $\delta$ ({\it cf.} Eq. \ref{M}) may give rise to a fake CPV signal even if the CP-odd phases $ \phi$ vanish, see \cite{ourreview,Afik:2021jjh} and discussion below. Thus, the construction of genuine CPV observables requires the joint consideration of a reaction and its CC counterpart \cite{Afik:2021jjh}, as described below.

It is useful to classify CP observables according to their transformation properties under ``naive time reversal'' ($T_N$), where $T_N$: time $\to -$time \cite{ourreview}. This classification is presented in Table~\ref{tab:tab1}. In particular, a $T_N$-odd CP-odd observable requires only a CP-odd phase, which could arise already at the tree-level, while a  $T_N$-even CP-odd observable needs also a non-vanishing CP-even phase from FSI (i.e., $\delta_i$ in Eq. \ref{M}), which is typically of a higher order. Thus, since $T_N$-odd CP-odd observables are sensitive to tree-level CPV, they 
are more likely to see
sizable CPV signals, in some cases reaching $\gtrsim 10\%$, see e.g., \cite{ourtthpaper,ourttZpaper,ourreview,ourPRL}. 

\begin{table}[htb]
\caption{General classification of CP-odd observables, ${\cal A}_{CP}$, under naive time reversal $T_N$, 
where $\Delta\delta$ and  $\Delta\phi$ denote  
the CP-even and CP-odd phases.
\label{tab:tab1}}
\begin{center}
\begin{tabular}{c|c|c}
 & $T_N$-odd (CP-odd) & $T_N$-odd (CP-even)  \tabularnewline
\hline \hline
\
CP-asymmetry & $A_{CP} \propto \cos\Delta\delta \sin\Delta\phi$ & $A_{CP} \propto \sin\Delta\delta \sin\Delta\phi$ \tabularnewline
\hline 
\
Required phases & only CP-odd & Both CP-odd \& CP-even \tabularnewline
\hline 
\
Sensitivity & tree-level CPV & CP-even phase from FSI  \tabularnewline
\
 &  & (higher order effect)  \tabularnewline
\hline \hline
\end{tabular}
\end{center}
\end{table}

Thus, in order to probe tree-level CPV, in the $2 \to 3$ process 
 $a b \to 1+ 2 +3$ and the CC reaction $\bar{a} \bar{b} \to \bar{1} + \bar{2} + \bar{3}$, we will use 
the following $T_N$-odd momenta correlations:
\begin{eqnarray}
\tp = \epsilon \left( p_a,p_1,p_2,p_3 \right)  ~~, ~~ 
\overline{\tp} = \epsilon \left( p_{\bar a},p_{\bar 1},p_{\bar 2},p_{\bar 3} \right) ~, \label{TP1}
\end{eqnarray}
which transform 
under $C$ and $CP$ as:
\begin{eqnarray}
    C(\tp) &=&  +\overline{\tp} ~,~ C(\overline{\tp}) =  + \tp~, \nonumber \\
    CP(\tp) &=&  -\overline{\tp} ~,~CP(\overline{\tp}) =  - \tp. 
\end{eqnarray}

Using the $\tp$'s in Eq.~\ref{TP1}, we then define the following 
$T_N$-odd (and also $P$-violating) asymmetries: 
\begin{eqnarray}
A_T &\equiv& \frac{N\left( \tp >0 \right) - N\left( \tp < 0 \right)}{N\left( \tp >0 \right) + N\left( \tp < 0 \right)} \label{AT1} ~, \\
\bar{A}_T &\equiv& \frac{N\left( -\overline{\tp} >0 \right) - N\left( -\overline{\tp} < 0 \right)}{N\left( -\overline{\tp} >0 \right) + N\left( -\overline{\tp} < 0 \right)} \label{AT2} ~,
\end{eqnarray}
where $N\left( \tp >0 \right)$ is the number of events for which ${\rm sign}(\tp) > 0$ is measured etc. 

The asymmetries $A_T$ and $\bar{A}_T$ are sensitive to the CP-odd phase {\it BUT} are not proper CP-asymmetries, since they are not eigenstate 
of CP, i.e., $CP(A_T) = \bar{A}_T$.  In particular, $A_T \propto \sin(\Delta\delta + \Delta\phi)$ and 
$\bar{A}_T \propto \sin(\Delta\delta - \Delta\phi)$ (see \cite{ourPRL}), so that in general $A_T \neq 0$ and/or $\bar{A}_T \neq 0$ could be generated 
without CPV, i.e., if $\Delta\phi =0$ and $\Delta\delta \neq 0$.
Therefore, in order to isolate the pure CPV effects, we need to combine the information from both $A_T$ and $\bar{A}_T$ as follows: 
\begin{eqnarray}
    {\cal A}_{CP} &=& \frac{1}{2} \left(A_T - \bar{A}_T \right) ~.  \label{ACP1}
\end{eqnarray}

As was shown in \cite{ourPRL}, when the initial state is not self-conjugate and its CC state has a different PDF, as can be the case at the LHC, i.e., PDF($ab$) $\neq$ PDF($\bar a \bar b$), then a 
``fake'' CP signal can still be generated in $A_{CP}$ in the presence of a CP-even phase, which is $\propto \cos\Delta\phi$. This is, however, not the case for the processes that we consider below as an example of tree-level CPV at the LHC (see Sect. \ref{CPV_LHC}). 

Finally, if the process under consideration is the same as the CC channel, i.e., $a b \to 1 +2 +3$ and $\bar{a} \bar{b} \to \bar{1} + \bar{2} + \bar{3}$ are the same (as is the case for example in $e^+ e^- \to t \bar t h$ that we will consider below in Sect. \ref{CPV_ILC}), then ${\cal A}_{CP} = A_T = \bar A_T$.

\section{CP-Violation from SM$\times$NP interference \label{tth_vertex}}

As explained in Sect. \ref{secIIA}, there is a single dimension 6 PTG operator, $\op_{t \phi}$, that may give rise to non-negligible CPV effects through interference with the SM in high-energy scattering processes; this operator modifies the $tth$ couplings and the interference effects  will not be proportional to a light quark mass. 

Given any model $M$, It is therefore convenient to parameterize the $tth$ generic interaction as follows: %
\begin{eqnarray}
    {\cal L}^{tth}_M = -  h \bar t \left(a_M + i b_M \gamma_5 \right) t ~, \label{tth_coup}
\end{eqnarray}
allowing for both scalar and pseudo-scalar couplings of the top quark to the Higgs; in particular, for the SM,
\begin{eqnarray}
    {\cal L}^{tth}_{SM} = y_{t} \, h\, \bar t t  ~~; ~~ y_{t}=\frac{ m_t}{v} ~, \label{ytSM}
\end{eqnarray}
so that $a_{SM}= m_t/v \sim 1/\sqrt{2}$ ($v \sim246\,$GeV is the Higgs VEV), and $b_{SM}=0$ (no pseudo-scalar coupling).

Considering now the SMEFT model ${\cal L}_{{\cal G}_t}$ of Eq.~\ref{eq:Gt}, which contains $\op_{t\phi}$, we find
\begin{eqnarray}
    {\cal L}^{tth}_{{\cal G}_t} = \frac{ m_t}v h\,  \bar tt  +  \frac{3v^2}{2\sqrt{2}\Lambda^2} h\, \left( \alpha_{t \phi}\bar t P_R t + {\rm H.c.} \right) ~,
\end{eqnarray}
so that 
\begin{eqnarray}
a_{{\cal G}_t} &=& a_{SM} + \frac3{2\sqrt{2}} \frac{v^2}{\Lambda^2} {\rm Re} \left( \alpha_{t \phi} \right) ~, \nonumber \\
b_{{\cal G}_t} &=& \frac3{2\sqrt{2}} \frac{v^2}{\Lambda^2} {\rm Im} \left( \alpha_{t \phi} \right) ~. \label{bGt}
\end{eqnarray}
Thus, with $\Lambda \sim 1$ TeV and a natural underlying heavy physics, i.e., ${\rm Re} \left( \alpha_{t \phi} \right) \sim 1$ and ${\rm Im} \left( \alpha_{t \phi} \right) \sim 1$, we have $a_{{\cal G}_t} \sim a_{SM} \sim 1/\sqrt{2}$ and a rather suppressed pseudo-scalar coupling 
$b_{{\cal G}_t} \sim \ 0.07$; in this case we  expect small CPV effects at high-energy scattering processes, as will be verified below.

On the other hand, multi-Higgs extensions of the SM could naturally give rise to a sizable pseudo-scalar coupling, of the order of $a_{SM}$. This is the case, for example, in the  two-Higgs doublet model  (2HDM) with CPV (generated by the  scalar potential, see e.g., \cite{ourreview}), where the lightest Higgs could have scalar and pseudo-scalar $tth$ couplings of $O(a_{SM})$, i.e., $a_{2HDM},b_{2HDM} \sim a_{SM} \sim 1/\sqrt{2}$; not a favorable scenario but not excluded experimentally   \cite{ATLAS1_th,ATLAS2_th,CMS1_th}. 

\bigskip

We will discuss below three processes which can potentially yield a CPV signal at tree-level, originating from NP in $tth$ interaction interfering with the SM at the tree-level: $e^+e^- \to t \bar t h$ at a future $e^+ e^-$ collide;r and two single-top -- Higgs associated production processes at the LHC, $pp \to thW$ and $pp \to thj$, where $j$ is a light-quark jet.  

\subsection{$e^+e^- \to t \bar t h$: CPV from tree-level 
 interference at a future $e^+ e^-$ collider \label{CPV_ILC}}

Tree-level CP-violation in $e^+e^- \to t \bar t h$ was first studied in \cite{ourtthpaper} (see also \cite{ourreview}), where  CPV in the $tth$ vertex was assumed to originate from a CP-odd phase in a 2HDM (similar tree-level CPV effects were also studied in \cite{ourttZpaper} for the case of $e^+e^- \to t \bar t Z$). Here we will generalize the study in \cite{ourtthpaper} to the EFT case, i.e., to any underlying heavy physics that can generate the CPV $tth$ interaction of Eq.~\ref{tth_coup}; in particular, discussing the expected CPV signal from the SMEFT operator $\op_{t \phi}$.  

\begin{figure}[htb]
  \centering
\includegraphics[width=0.8\textwidth]{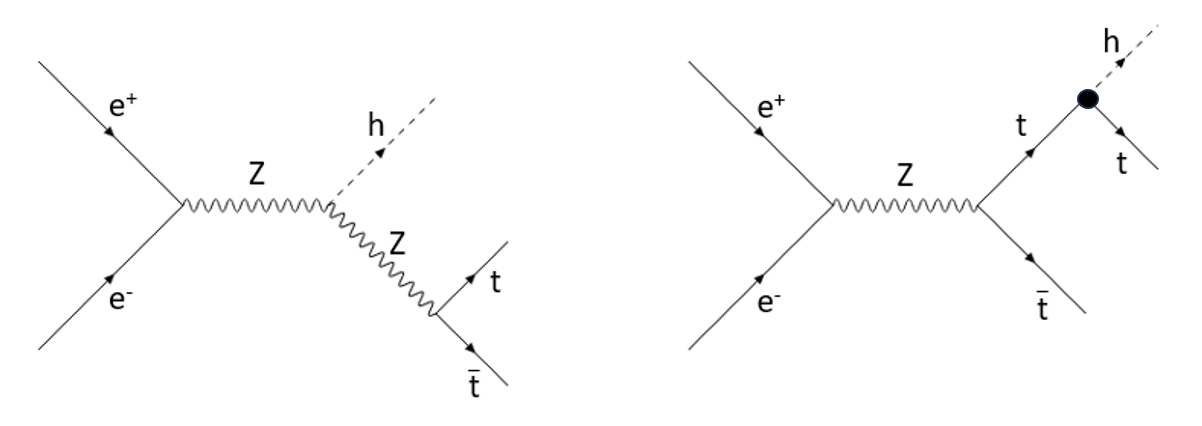}
\caption{Representative Feynman diagrams for the lowest order SM and NP contributions (marked by a heavy dot) to $e^+ e^- \to t \bar t h$ at a future $e^+ e^-$ collider. Tree-level CP-violation arises from interference between these two diagrams. See also text.}
  \label{fig:tth}
\end{figure}
\begin{figure}[htb]
  \centering
\includegraphics[width=0.45\textwidth]{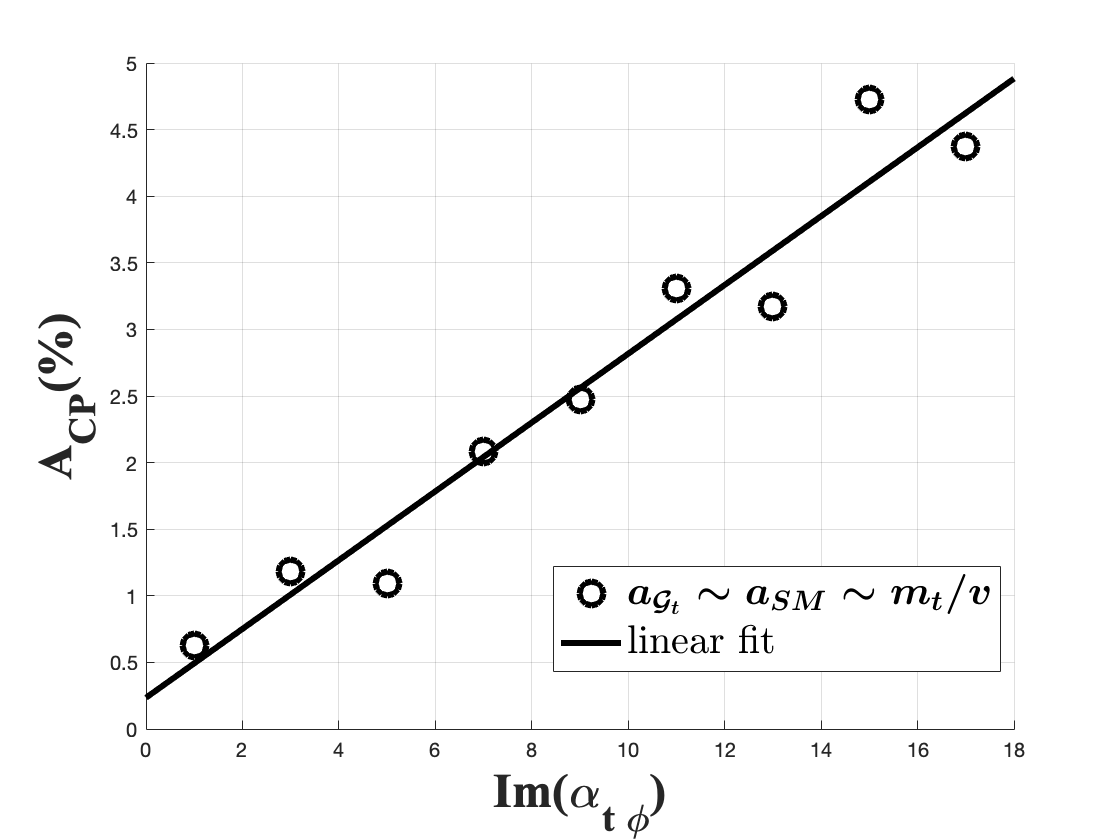}
\includegraphics[width=0.45\textwidth]{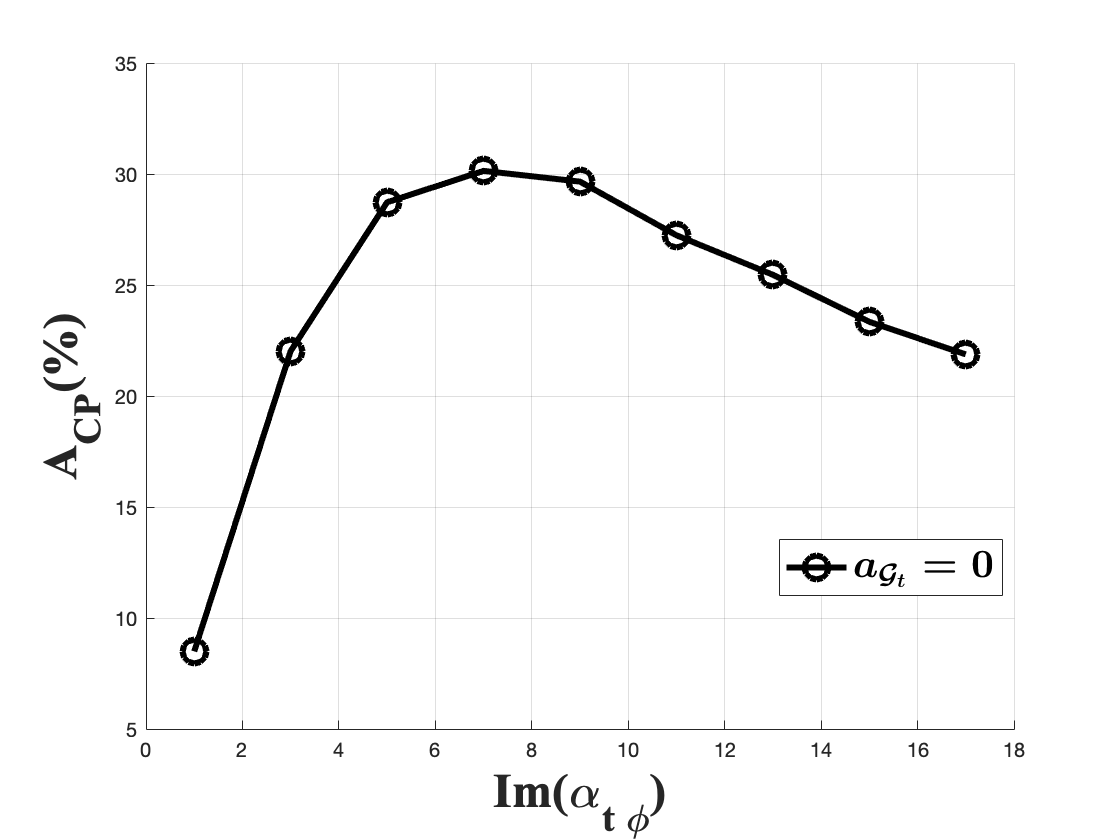}
\caption{The CP-aymmetry as a function of the imaginary element of $\alpha_{t \phi}$, for $\Lambda=1$ TeV. Left plot: the case of $a_{{\cal G}_t} \sim  a_{SM} = m_t/v$ and $b_{{\cal G}_t} \sim \frac{v^2}{\Lambda^2} \cdot {\rm Im} (\alpha_{t\phi})$. Right plot: the case where $a_{{\cal G}_t} =0$ and $b_{{\cal G}_t} \sim \frac{v^2}{\Lambda^2} \cdot {\rm Im} (\alpha_{t\phi})$. See text for more details.}
  \label{fig:tth_ACP}
\end{figure}

In this process the tree-level CPV effect arises 
from interference between the diagrams in
Fig.~\ref{fig:tth}. To leading order in the SMEFT expansion, the CPV part of the differential cross-section (see Eq.~\ref{sigCPV1}) is (see also \cite{ourtthpaper,ourreview}):
\begin{eqnarray}
    d\sigma_{CPV}^{tth} \propto m_t \cdot c_{hZZ} \cdot b_{{\cal G}_t} \cdot \epsilon \left(p_{e^-},p_{e^+},p_{t},p_{\bar t} \right) ~,\label{dsigCPV3l}
    \end{eqnarray}
where $b_{{\cal G}_t}$ is the pseudo-scalar $tth$ coupling 
from the CPV SMEFT$_t$ sector (see Eq.~\ref{bGt}),
$c_{hZZ}$ represents the $hZZ$ coupling:
\begin{eqnarray}
    {\cal L}_{hZZ} = m_Z  \cdot c_{hZZ} h\,  Z^\mu Z_\mu ~,
\end{eqnarray}
and $c_{hZZ} = g_W m_Z/m_W$ in the SM. The CPC part of the differential cross-sections, $d\sigma_{CPV}^{tth}$, contains terms proportional to $a_{{\cal G}_t}^2$, $b_{{\cal G}_t}^2$, $c_{hZZ}^2$ and $a_{{\cal G}_t} \cdot c_{hZZ}$.

%
%

In Fig.~\ref{fig:tth_ACP} we plot the CP-asymmetry ${\cal A}_{CP}$ of Eq.~\ref{ACP1}, as a function of ${\rm Im}(\alpha_{t \phi})$, the imaginary part of the Wilson coefficient of the operator $\op_{t \phi}$. For natural and weakly-interacting NP we expect $\alpha_{t \phi} \sim 1$ (for both the real and imaginary parts of $\alpha_{t \phi}$), so that 
for $\Lambda^2 \gg  v^2$, we have for the scalar and pseudoscalar $tth$ couplings in Eq.~\ref{bGt}:
\begin{eqnarray}
a_{{\cal G}_t} \sim a_{SM} = O(1) ~~ ,~~
b_{{\cal G}_t} \sim O\left(\frac{v^2}{\Lambda^2}\right) ~.
\label{bGt2}
\end{eqnarray}
whence we expect (recall that ${\cal A}_{CP} \propto d\sigma_{CPV}^{tth}/d\sigma_{CPC}^{tth}$, see Eq.~\ref{ACP_gen}):
\begin{eqnarray}
    {\cal A}_{CP} \sim O\left(\frac{v^2}{\Lambda^2}\right) \sim O(1\%) ~, \label{ACPtth_nat}
\end{eqnarray}
as indeed shown  on the left panel in Fig.~\ref{fig:tth_ACP} for 
the case ${\rm Im} (\alpha_{t\phi}) \sim 1$. 

On the right panel of Fig.~\ref{fig:tth_ACP} we consider the case where  the scalar $tth$ coupling vanishes, $a_{{\cal G}_t} =0$ (these results also apply to the more general case where $a_{{\cal G}_t} \ll b_{{\cal G}_t}$). This is an unnatural choice for the parameter space of the SMEFT framework, 
but it could apply to model dependent scenarios with new particles of sub-TeV  masses, such as the 2HDM, as was investigated in \cite{ourtthpaper} (see also \cite{ourreview}). We see from Fig.~\ref{fig:tth_ACP} that, in such a case, a CP-asymmetry of up to ${\cal A}_{CP} \sim 10-30\%$ is possible in $e^+ e^- \to t \bar t h$, 
for ${\rm Im} (\alpha_{t\phi}) \sim 1 - 10$ (note that 
${\rm Im} (\alpha_{t\phi}) \sim 20$ corresponds to a pseudo-scalar $tth$ coupling  $b_{{\cal G}_t} \sim 1$).

\subsection{Tree-level CPV from interference in top-Higgs associated production at the LHC \label{CPV_LHC}}

We consider here the expected sensitivity to potential CPV signals in scattering processes at the LHC, focusing again on the leading CPV effects within the SMEFT framework; namely, on the effects expected from the interference between the PTG operator $\op_{t \phi}$ and the SM. As before, a CP-odd phase $\propto {\rm Im}(\alpha_{t\phi})$ is assumed to be present in the modified $tth$ interaction.

Let us first address the possibility of detecting CPV in $pp \to t \bar t h$, which is the LHC analog of the processes $e^+ e^- \to t \bar t h$ described above. In contrast  to $e^+ e^-$ colliders,  $t \bar t h$ production at the LHC is dominated by the QCD gluon-gluon fusion process $gg \to t \bar t h$, which is CP-invariant at leading order even in the presence of the modified CPV $tth$ interaction of Eq.~\ref{tth_coup}. As a consequence, the CPV asymmetry in $pp \to t \bar t h$ is too small to be detected at the LHC. Specifically,  as in the case of $e^+ e^- \to t \bar t h$ described above, CPV in  $pp \to t \bar t h$ is generated by
interference of the $Z$-boson mediated electroweak (EW) processes  $q \bar q \to Z \to t \bar t h$ where the Higgs is radiated from the $Z$-boson, with the diagrams where the Higgs is emitted from the top and/or anti-top quarks in this process, so the asymmetry  is doubly suppressed: {\it(i)} by the ratio between the PDF's of the initial $q \bar q$ and $gg$ states, i.e., PDF($q \bar q$)/PDF($gg$); and {\it(ii)} by a ratio of EW/QCD couplings. Unsurpisingly, a recent search by ATLAS \cite{ATLAS2_th}, using dedicated CP-sensitive observables in $pp \to t \bar t h$, found no evidence for CPV. 

\begin{figure}[htb]
  \centering
\includegraphics[width=0.8\textwidth]{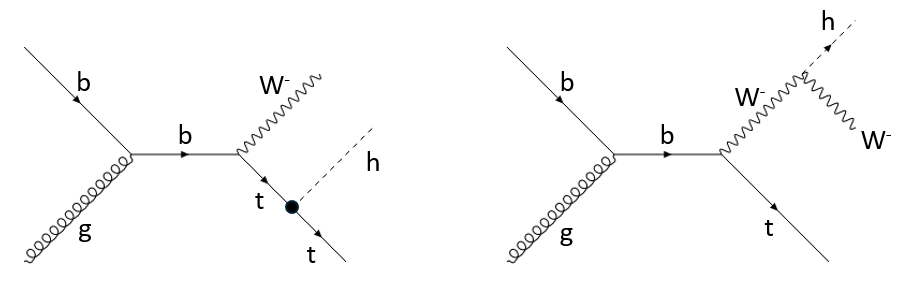}
\includegraphics[width=0.6\textwidth]{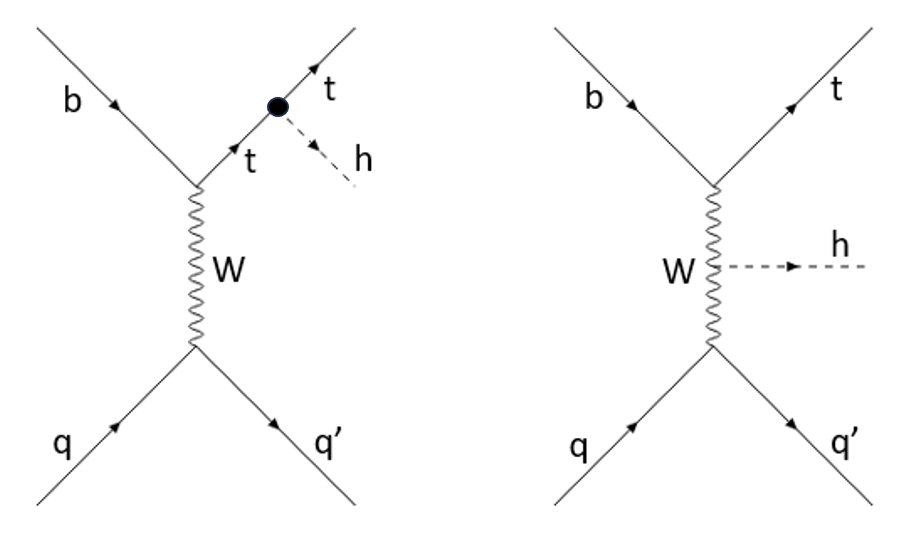}
\caption{Representative Feynman diagrams for the lowest order SM and NP contributions (marked by a heavy dot) to the $pp \to t h W$ (upper diagrams) and $pp \to t h j$ (lower diagrams) processes at the LHC. In each case, tree-level CP-violation arises from interference between the two diagrams. See also text.}
  \label{fig:tWh}
\end{figure}

Thus, to search for tree-level CPV effects at the LHC from the PTG operator $\op_{t \phi}$, 
one has to look for alternative scattering processes involving top-Higgs associated production.  This  leads us to consider the following two single-top plus Higss production channels: 
\begin{itemize}
    \item \underline{\boldmath{$thW$} \bf{associated production}:} $pp \to thW^-$ and the CC process $pp \to \bar thW^+$, which are mediated by the gluon-$b$ fusion processes $gb \to t h W^-$ and $g \bar b \to \bar t h W^+$, respectively (see Fig.~\ref{fig:tWh}). 
    \item \underline{\boldmath{$thj$} \bf{associated production}:} $pp \to thj$ and the CC process $pp \to \bar th j$, where $j$ is a light quark or anti-quark jet, which are mediated by the $q$-$b$ fusion processes $qb \to t h j$ and $\bar q \bar b \to \bar t h j$, respectively, where $q$ is either an up-quark or down anti-quark; the leading effect is from $u,d$ of the first generation (see Fig.~\ref{fig:tWh}). 
\end{itemize}

Indeed, these top-Higgs associated production processes are 
at the focus of the ATLAS \cite{ATLAS1_th,ATLAS2_th} and CMS \cite{CMS:2020cga,CMS1_th} searches, due to their noticeable sensitivity to NP in general \cite{tH3,Degrande:2018fog,Jose_thj} and specifically to the $tth$ Yukawa interaction \cite{tH1,tH2,tH3,tH4,tH5,1607.05862,Barger:2019ccj,2007.08542}. As shown below, they may also exhibit tree-level CPV from a CP-odd phase of the PTG operator $\op_{t \phi}$, via the interference between the diagram where the Higgs is radiated from the $W$-boson and the diagram where it is emitted from the top-quark.

It should also be noted that the total cross-section for each of the $tH$ associated production processes is itself very sensitive to the interference between these two diagrams  \cite{tH2,tH3,tH4,1607.05862,Barger:2019ccj,2007.08542}. Therefore, regardless of CPV, these two $th$ production channels provide direct access to the size and sign of the $tth$ Yukawa interaction, as their cross-sections can vary by more than an order of magnitude depending on the size and sign of the $tth$ Yukawa coupling $y_t$ (or equivalently, on the size and sign of $a_M$ in Eq.~\ref{tth_coup}). We will show below that the CP asymmetry for these two $th$ production channels is also sensitive to the sign (and size) of the scalar $tth$ Yukawa coupling $y_t$.\footnote{Due to our focus on CPV we only consider the effects of the SMEFT operator $\op_{t\phi}$. A complete calculation of possible NP effects on other observables associated with these reactions required the inclusion of a number of other operators, see, e.g. \cite{Jose_thj}.} 

In order to asses the maximal size of a CPV effect in $th$ associated production at the LHC due to a CP-odd phase in the $tth$ Yukawa coupling from the operator $\op_{t \phi}$, i.e., the pseudo-scalar coupling $b_{{\cal G}_t}$ in Eq.~\ref{bGt},  we will perform below a "sterile" analysis. In particular, we will not consider here the SM backgrounds to these processes (see e.g., \cite{ATLAS1_th,ATLAS2_th,CMS1_th}) and issues concerning the separation 
of the $th$ signals 
from (the dominant) $pp \to t \bar t h$ process; in particular for the $thW$ case \cite{1607.05862,2007.08542}. Also, we will not include 
the decays of the top-quark, of the Higgs and of the $W-boson$ and the corresponding detection efficiencies.
We emphasize, that such 
issues are expected to further dilute the expected CPV effect in $th$ associated production reported below, and we leave them to a future work which will be dedicated to the investigation of CPV in single top + Higgs associated production at the LHC.

Let us now turn to the study of CPV effect in the aforementioned $th$ associated production channels at the LHC. 
In analogy to the $e^+e^- \to t \bar t h$ case described in the previous section, the CPV parts of the differential hard cross-sections $g b \to thW$ and $qb \to thj$ are:
\begin{eqnarray}
    d\sigma_{CPV}^{thW,j} &\propto& m_t \cdot c_{hWW} \cdot b_{{\cal G}_t} \cdot \epsilon \left(p_{b},p_{t},p_{h},p_{W^-,j}\right) ~, 
    \label{dsigCPVtWh}
    \end{eqnarray}
where $b_{{\cal G}_t}$ is the pseudo-scalar $tth$ coupling in the SMEFT$_t$ framework (see Eq.~\ref{bGt}) and $c_{hWW}$ represents the SM $hWW$ coupling.  
For the CC channels, $g \bar b \to \bar t W^+ h$ and $\bar q \bar b \to \bar thj$, replace $\epsilon \left(p_{b},p_{t},p_{h},p_{W^-}\right) \to \epsilon \left(p_{\bar b},p_{\bar t},p_{h},p_{W^+} \right)$ and 
$\epsilon \left(p_{b},p_{t},p_{h},p_{j}\right) \to \epsilon \left(p_{\bar b},p_{\bar t},p_{h},p_{\bar j} \right)$, respectively. 
Here also, the corresponding CPC part of the differential cross-sections contains terms proportional to $a_{{\cal G}_t}^2$, $b_{{\cal G}_t}^2$, $c_{hWW}^2$ and $a_{{\cal G}_t} \cdot c_{hWW}$.

\begin{table}[htb]
\begin{center}
\caption{The expected CP asymmetries in $pp \to thW$ and $pp \to thj$ events, for $\Lambda=1$ TeV, ${\rm Im}(\alpha_{t \phi}) = 1$ and a $tth$ Yukawa coupling $y_t = 0,\, \pm m_t/v$ (see Eq.~\ref{ytSM}). A selection cut was used on the c.m. energy $\sqrt{\hat s} < \Lambda$; see also text. \label{tab:data1}}
\begin{tabular}{c|c|c|c|c|c|c|}
 & \multicolumn{3}{c|}{$pp \to thW$} & \multicolumn{3}{c|}{$pp \to thj$} 
\tabularnewline
\cline{2-7} 
\
 & $ ~~~y_t=0 ~~~$ & $y_t=m_t/v$ & $y_t=-m_t/v$ & $~~~ y_t=0 ~~~$ & $y_t=m_t/v$ & $y_t=-m_t/v$
\tabularnewline
\hline 
\hline
\
${\cal A}_{CP}$ & 0.2\% & 1.3\% & 0.1\% & 0.8\% & 1.7\% & 0.05\% 
\end{tabular}
\end{center}
\end{table}
\begin{table}[htb]
\begin{center}
\caption{The cross-sections $\sigma(pp \to thW) = \sigma(pp \to thW^-)=\sigma(pp \to \bar{t} h W^+)$, $\sigma(pp \to thj)$ and $\sigma(pp \to \bar{t}hj)$, at LO with $\Lambda=1$ TeV, ${\rm Im}(\alpha_{t \phi}) = 1$ and a selection $\sqrt{\hat s} < \Lambda$, for a $tth$ Yukawa coupling $y_t = 0, \, \pm m_t/v$ (see Eq.~\ref{ytSM}). \label{tab:data12}}
\begin{tabular}{c|c|c|c|}
 & $~~~~y_t=0~~~~$ & $y_t=m_t/v$ & $y_t=-m_t/v$ 
\tabularnewline
\hline 
\hline
\
$\sigma(pp \to thW)$ [fb] & 10.1 & 6.9 & 42.9 
\tabularnewline
\hline
$\sigma(pp \to thj)$ [fb] & 71.8 & 18.0 & 268.2 
\tabularnewline
\hline
$\sigma(pp \to \bar{t}hj)$ [fb] & 39.5 & 8.5 & 143.1 
\end{tabular}
\end{center}
\end{table}

In Table \ref{tab:data1} we list the expected CP asymmetry ${\cal A}_{CP}$ for the $thW$ and $thj$ signals at the LHC, for $\Lambda= 1$ TeV, ${\rm Im}(\alpha_{t \phi}) = 1$ and for  three values of the top-quark Yukawa coupling 
$y_t = 0,\,\pm m_t/v$. In Table \ref{tab:data12} we list the corresponding leading order (LO) cross-sections. 
An upper selection cut  $\sqrt{\hat s} < \Lambda$, i.e., on the center of mass energy of the hard processes, was applied on all event samples to ensure the validity of the EFT prescription.

\begin{figure}[htb]
  \centering
\includegraphics[width=0.45\textwidth]{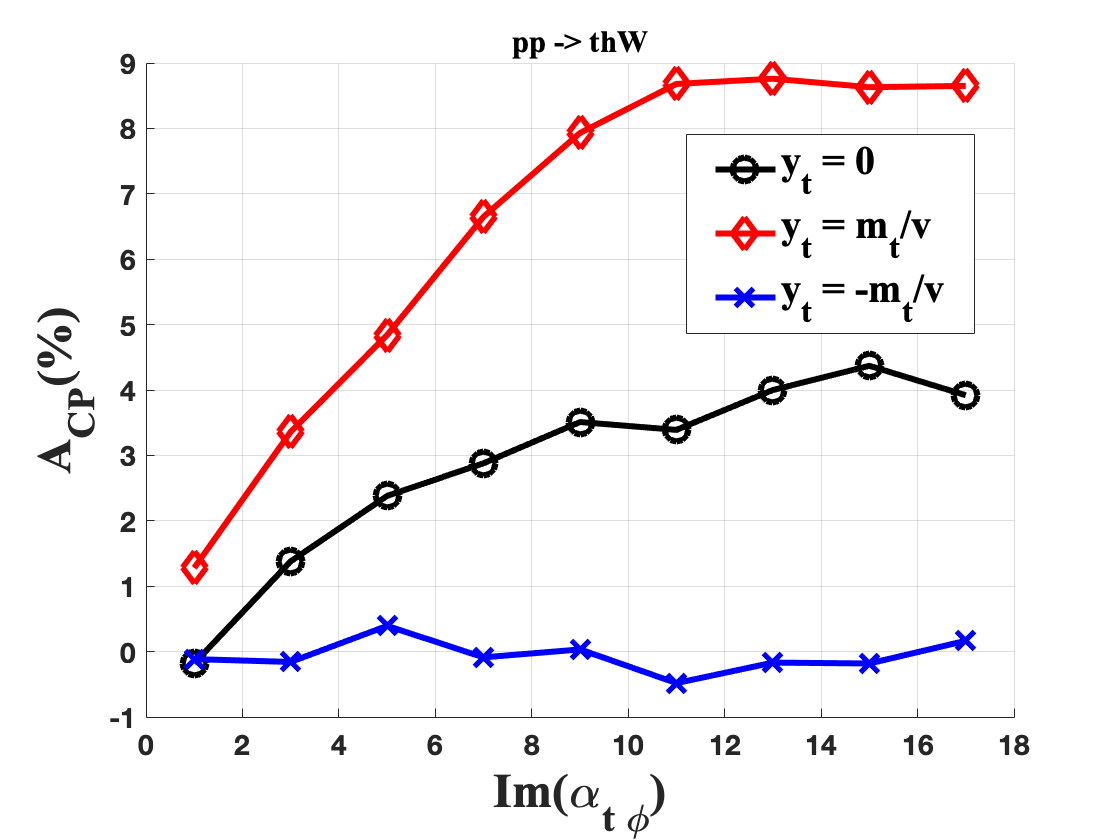}
\includegraphics[width=0.45\textwidth]{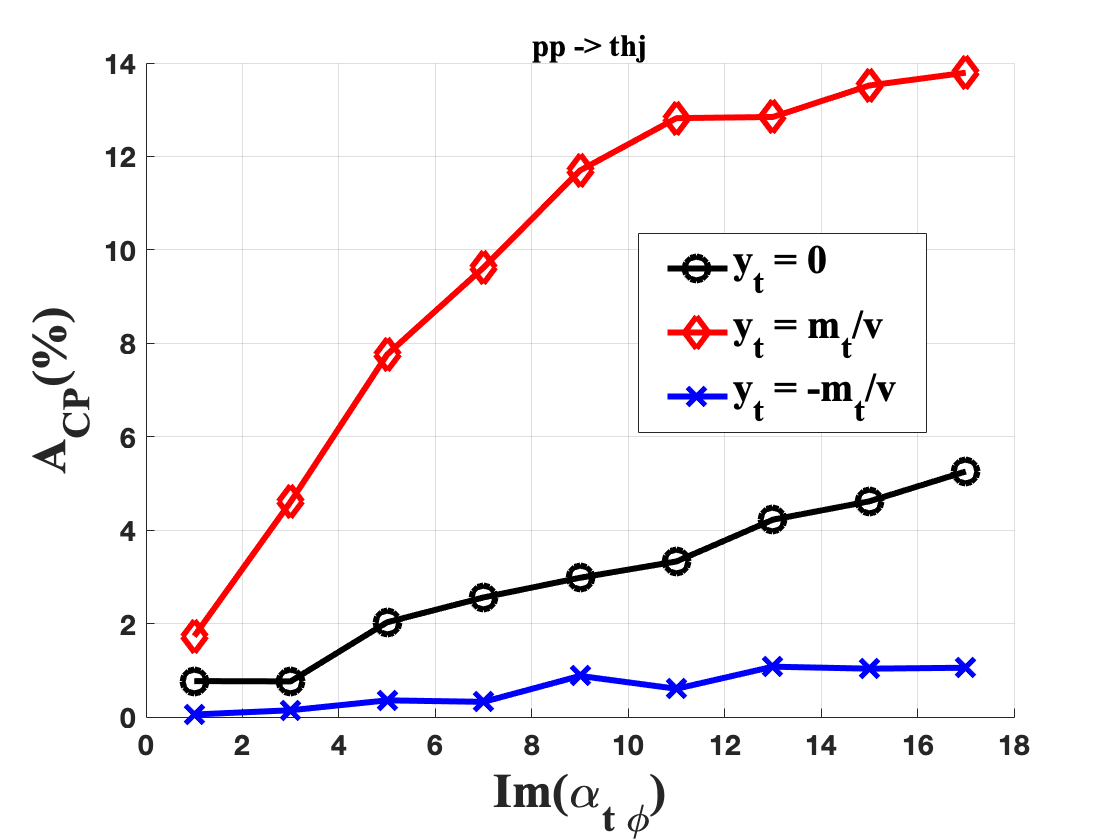}
\caption{The expected CP-asymmetries in $thW$ (left) and $thj$ (right) associated production at the LHC, as a function of the imaginary element of $\alpha_{t \phi}$, for $\Lambda=1$ TeV and for $y_t=0,\, \pm m_t/v$, respectively, where $y_t$ is the $tth$ Yukawa coupling (see Eq.~\ref{ytSM}). See also text.}
  \label{fig:th_ACP}
\end{figure}

The behaviour of ${\cal A}_{CP}$ for these two processes as a function of ${\rm Im} (\alpha_{t\phi})$, for $y_t=0,\,\pm m_t/v $, is depicted  in Fig.~\ref{fig:th_ACP}.\footnote{The $th$ event samples were generated using {\sc MadGraph5\_aMC@NLO}~\cite{madgraph5} at LO parton-level and with the SMEFTsim model of~\cite{SMEFTsim1,SMEFTsim2} for the EFT framework. The 5-flavor scheme was used to generate all samples, with the ${\tt NNPDF30\_lo\_as\_0130}$ PDF set~\cite{Ball:2014uwa} and the default {\sc MadGraph5\_aMC@NLO} LO dynamical scale.
} As described in section \ref{subsec:CPV1}, the CP-asymmetry ${\cal A}_{CP}$ is constructed out of the $T_N$-odd observables, $A_T$, ${\bar A}_T$ of Eqs.~\ref{AT1} and \ref{AT2}.

We observe three notable properties of the expected ${\cal A}_{CP}$ in $th$ associated production at the LHC:
\begin{itemize}
    \item As expected (see above), for natural Wilson coefficients (${\rm Im} (\alpha_{t\phi}), {\rm Re} (\alpha_{t\phi}) \sim  1$),  the CP asymmetry is, at best, of $O(1\%)$. A naive estimate of the detection sensitivity to such an asymmetry, based on statistics only (see Eq.~\ref{NSD}), shows that at least $O(10^5)$ $pp \to thW$ events would be required for a $3 \sigma$ signal; this is not within the reach of the LHC. 
\item A sizable CP asymmetry, of ${\cal A}_{CP} \sim 10\%$, can arise only for ${\rm Im} (\alpha_{t\phi}) \gsim 10$, which is unnaturally large within the EFT approach. However, as mentioned earlier, this case represents model-dependent scenarios of sub-TeV NP, e.g., multi-Higgs models, and not of multi-TeV (decoupling) NP scenarios.  
\item The strong interference effects noted previously, which depends on the size and sign of the top-quark Yukawa coupling, is clearly manifested also in ${\cal A}_{CP}$. In particular, the asymmetry is maximal in both the $thW$ and $thj$ channels, if $y_t =+m_t/v$ and it is strongly suppressed for $y_t = - m_t/v$. It is interesting to note that this behaviour of ${\cal A}_{CP}(y_t)$ is  opposite to that of the cross-section, 
see Table \ref{tab:data12} and \cite{tH2,tH3,tH4,1607.05862}; the reason is that the total (CP-conserving) cross-section appears in the denominator of ${\cal A}_{CP}$, see Eqs.~\ref{AT1}-\ref{ACP1}. 
\end{itemize}

\section{Tree-level CPV in tri-lepton events from pure NP dynamics \label{FCNP_CPV}}

In this section we will describe a case of a potential large tree-level CPV effect, of $O(10\%)$, from natural underlying NP which has $O(1)$ Wilson coefficients in the EFT prescription; this occurs within the TLCPV-III scenario defined in Sect. \ref{CPVtypes}, when there is no (irreducible) SM tree-level contribution to the cross section. The results presented below are based on
\cite{ourPRL}, where a more detailed analysis can be found. 

Specifically, we consider the tri-lepton production at the LHC 
$a b \to \ell^{\prime -} \ell^+ \ell^- + X$ and the CC channel 
$\bar{a} \bar{b} \to \ell^{\prime +} \ell^- \ell^+ + \bar X$. 
The underlying NP responsible for CPV in these tri-lepton signals at the LHC is assumed to correspond to the following two  $t u \ell \ell$ 4-Fermi effective operators \cite{ourPRL}:\footnote{We assume  that the two operators have a common underlying scale, experimentally $\Lambda \gsim 1$~TeV, see \cite{Afik:2021jjh}.}
\begin{eqnarray}
\op_S = \left(\bar\ell_R \ell_R \right) \left( \bar t_R u_R \right) ~~,~~ \op_T = \left(\bar\ell_R \sigma_{\mu \nu} \ell_R \right) \left( \bar t_R \sigma_{\mu \nu} u_R \right) \,; \quad \ell = e,\,\mu \label{eq:OST} 
\end{eqnarray}
and similarly for the $t c \ell \ell$ 4-Fermi terms. The relevant Lagrangian is then
\begin{eqnarray}
{\cal L} = {\cal L}_{SM} + \frac{1}{\Lambda^2} \left[ \left( \alpha_S \op_S + \alpha_T \op_T \right) + \text{H.c.} \right] ~, \label{Ldim6}
\end{eqnarray}

\begin{figure}[htb]
  \centering
\includegraphics[width=0.75\textwidth]{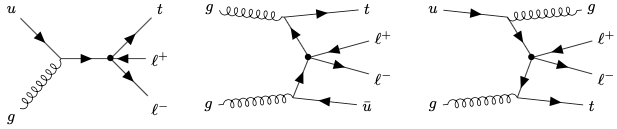}
\caption{Representative Feynman diagrams for the lowest order single top-quark + di-lepton production channels $pp \to t \ell^+ \ell^-$ and $pp \to t \ell^+ \ell^- + j$ ($j$ is a light jet), via the $t u \ell^+ \ell^-$ 4-Fermi interaction (marked by a heavy dot).}
  \label{fig:Feynman}
\end{figure}

These two operators (which correspond to the 4-Fermi operators $\op_{\ell e q u}^{(1,3)}$ in Eq.~\ref{nonH}) are not Hermitian, so, in general, their Wilson coefficients will contain CP-odd phases. Contributions from $\op_{S,\,T}$   can then interfere with each other (even if not with the SM), generating CPV effects. This indeed occurs in tri-lepton events at the LHC in the single-top production channels (see Fig.~\ref{fig:Feynman}):
\begin{eqnarray}
ug, gg \to \ell^+ \ell^- t,~ \ell^+ \ell^- t +j  \quad (j={\rm light~jet}) \label{eq:tll} 
\end{eqnarray}
and the CC counterparts (i.e., with $\bar t$ in the final state).

Thus, considering CPV in the single-top production channels in Eq.~\ref{eq:tll},  followed by the top decay to a charged lepton, we will focus, as an example, on the $e \mu \mu$ final state:\footnote{The processes in Eq.~\ref{eq:tll}, followed by the top decay $t \to Wb \to \ell^\prime \nu_{\ell^\prime} b$, have interesting implications also for generic NP searches of new heavy states around the TeV scale, which can generate the top-leptons 4-Fermi contact terms $\op_{S,\,T}$ \cite{Afik:2021jjh}.}
\begin{eqnarray}
pp \to  t \mu^+ \mu^- +X \to e^+ \mu^+ \mu^- +X ~, \nonumber \\
pp \to  \bar{t} \mu^+ \mu^- +\bar{X} \to e^- \mu^- \mu^+ + \bar{X} ~, \label{eq:emumu} 
\end{eqnarray}
where the NP and CPV phases arise from the $tu \mu \mu$ and/or $tc \mu \mu$ 4-Fermi interactions in Eq.~\ref{Ldim6}.  

As was shown in \cite{ourPRL}, the dominant SM background to the inclusive tri-lepton production channels of Eq.~\ref{eq:emumu} 
arises from $WZ$ production, $pp \to WZ$, followed by the $W$ and $Z$ decays to charged leptons. 
The  SM contributions to $pp \to  e^+ \mu^+ \mu^- +X$ from other channels such as
$pp \to t \bar t, t \bar{t} V, tVV, V+jets$ ($V=Z,W$)  are much smaller \cite{ourPRL}. 
In particular, 
there is no interference between the $tu \mu\mu$ NP diagrams (see Fig.~\ref{fig:Feynman}) and the SM contributions to the tri-leptons signal, 
so that the cross-section is of the form: 
\begin{equation}
\sigma_{e\mu\mu}(m_{\mu\mu}^{min}) = \sigma^{SM}_{e\mu\mu}(m_{\mu\mu}^{min}) + \frac{\alpha^2}{\Lambda^4} \sigma^{NP}_{e\mu\mu}(m_{\mu\mu}^{min})~,
\end{equation}
where $\alpha^2$ stands for all the NP$\times$NP coefficients (i.e., proportional to $|\alpha_S|^2$, $|\alpha_T|^2$ and $\alpha_S^* \alpha_T$) and we have defined the $m_{\mu\mu}^{min}$-dependent cumulative cross-section, 
selecting events with $m_{\mu\mu} > m_{\mu\mu}^{min}$: 
\begin{eqnarray}
\sigma_{e\mu\mu}(m_{\mu\mu}^{min}) \equiv \sigma_{e\mu\mu}(m_{\mu\mu} > m_{\mu\mu}^{min}) = 
\int_{m_{\mu\mu} \geq m_{\mu\mu}^{min}} d m_{\mu\mu} \frac{d \sigma_{e\mu\mu}}{dm_{\mu\mu}} ~. \label{cum} 
\label{CCSX}
\end{eqnarray}

The invariant mass of the two muons in the final state, $m_{\mu\mu}$, turns out to be a useful discriminating parameter;\footnote{In the more general case of $pp \to \ell^{\prime \pm} \ell^+ \ell^- +X$, 
we have $m_{\mu\mu} \to m_{\ell \ell}$ and $m_{\ell \ell}$ would be the invariant mass of the "none-top" opposite sign same-flavor (OSSF) di-leptons from the underlying hard process, 
i.e., of the di-leptons produced from the $tt \ell \ell$ vertex and not from the top-quark decays.} the CP asymmetries are sensitive to the di-muons invariant mass, since 
the SM (which is dominant at low $m_{\mu \mu}$) contributes to the denominators while the CPV NP term
(which is dominant for high $m_{\mu \mu}$) contributes to the numerators. 

The CPV in this case is, therefore, a pure NP effect, since it arises from the imaginary part of the interference between the  dimension 6 scalar and  tensor operators, provided at least one of the corresponding Wilson coefficients  is complex. In particular, the numerator of the CP-asymmetry ${\cal A}_{CP}$ (and of the $T_N$-odd asymmetries $A_T$ and $\bar{A}_T$) is proportional to the CPV part of the 
cross-section for $p p \to t \mu^+ \mu^- \to e^+ \mu^+ \mu^- +X$:
\begin{eqnarray}
   d\sigma_{CPV} \propto \epsilon \left( p_{u_i},p_{e^+},p_{\mu^+},p_{\mu^-} \right) \cdot {\rm Im} 
   \left( \alpha_S \alpha_T^\star \right)~, \label{sigCPV}
\end{eqnarray}
and similarly for the CC channel by replacing 
$\epsilon \left( p_{u_i},p_{e^+},p_{\mu^+},p_{\mu^-} \right)$ with $\epsilon \left( p_{\bar{u}_i},p_{e^-},p_{\mu^-},p_{\mu^+} \right)$. 

We present below results for $\vert \alpha_S \vert = 1$, $\vert \alpha_T \vert = 0.25$ (which is motivated by the matching of the EFT framework to  leptoquarks models, see \cite{ourPRL}), with a maximal CP-odd phase for the $t u \mu \mu$ and $t c \mu \mu$ operators, so that the CPV coupling in Eq.~\ref{sigCPV} is set to:
\begin{equation}
    {\rm Im} \left( \alpha_S \cdot \alpha_T^\star \right) = 0.25 ~. \label{CPVvalue}
\end{equation}

\begin{figure}[htb]
  \centering
\includegraphics[width=0.50\textwidth]{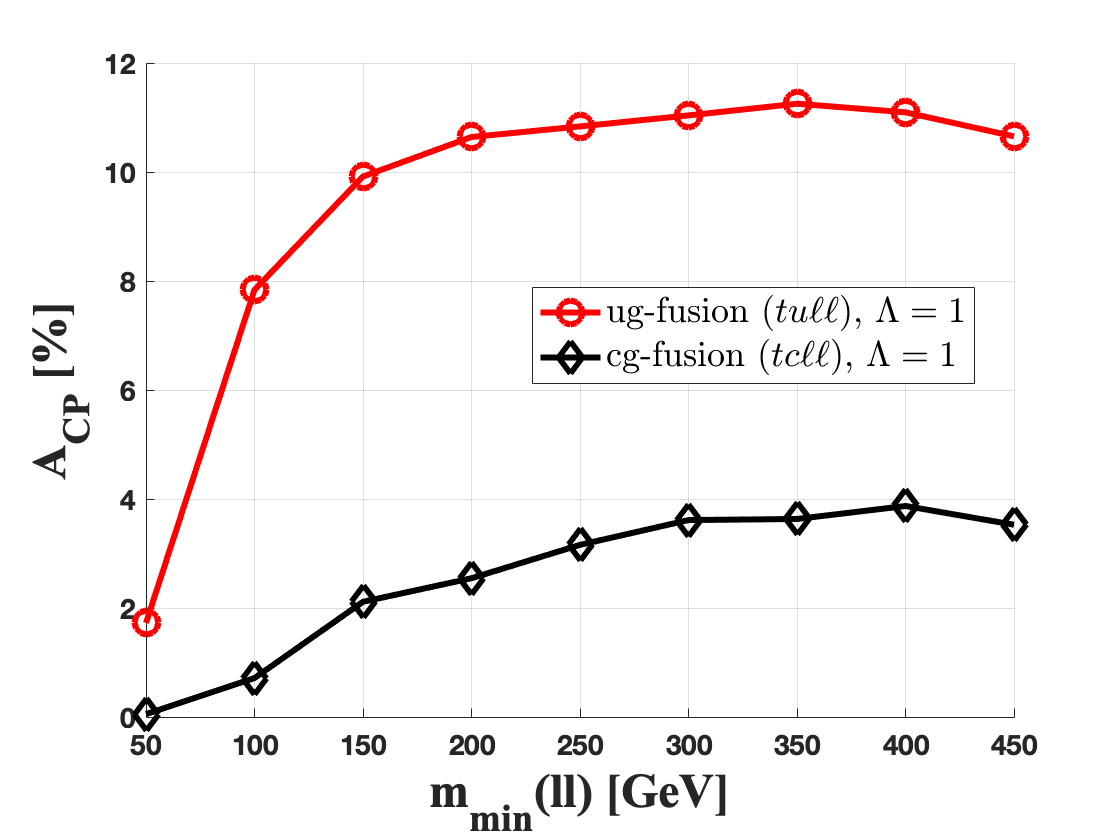}
\caption{The expected CP-asymmetry ${\cal A}_{CP}$, as a function of the invariant mass cut on the same-flavor di-leptons, see text. 
We are only showing central values, for more details see \cite{ourPRL}.}
\label{fig:ACP}
\end{figure}

In Table~\ref{tab:data2}, we list the resulting CPV and $T_N$-odd asymmetries for $m^{min}_{\ell \ell}=400$~GeV, and in Fig.~\ref{fig:ACP} we show the dependence of ${\cal A}_{CP}$ on $m^{min}_{\ell \ell}$.  Results are shown for both the $ug$-fusion and $cg$-fusion production channels, for $\Lambda=1,2$ TeV,  and include SM effects background from $pp \to ZW^{\pm} +X$. 
\begin{table}[htb]
\begin{center}
\caption{The expected $T_N$-odd and CP asymmetries in tri-lepton events, $pp \to \ell^{\prime \pm} \ell^+ \ell^- +X$, see text. Table taken from \cite{ourPRL}}
\begin{tabular}{c|c|c}
 & $ug$-fusion: $\Lambda=1(2)$~TeV & $cg$-fusion: $\Lambda=1(2)$~TeV 
\tabularnewline
\hline \hline
\
${\cal A}_{CP}$ & 11.1(7.9)\% & 3.9(0.7)\% \tabularnewline
\hline 
\
$A_{T}$ & 16.4(13.5)\% & 3.1(0.5)\% \tabularnewline
\hline 
\
$\bar{A}_T$ & -5.8(-2.3)\% & -4.7(-1.0)\% \tabularnewline
\hline \hline
\end{tabular}
\end{center}
\end{table}

\section{Summary}

We have systematically analysed the discovery potential of CP-violation searches in scattering processes at TeV-scale colliders, parametrizing the underlying heavy NP in an effective field theory framework and using the SMEFT basis for the higher dimensional operators. 

We have based our methodology on naturality arguments and 
possible structures of the potential heavy physics that underlies the SM, which is assumed to be weakly interacting and decoupling.   
We have then considered the phenomenological implications of the CP-violating sector of the SMEFT framework in three limiting cases: (i) an enhanced U(3)$^5$ flavour symmetry that the SM possesses when all fermion masses are set to zero, (ii) a reduced U(3)$^4$ flavour symmetry where top-quark mass effects are included and (iii) no flavour symmetry. 

In particular, we have 
spelled out guiding principles for the hunt of CPV in high-energy scattering processes, finding that if flavor changing interactions are suppressed (or absent) in the underlying heavy physics, then the leading CPV signals are expected from a single dim.6 operator that involves the top-quark and Higgs fields, $\op_{t\phi} = \phi^\dagger \phi  \left(\bar q_3 t \right) \tilde{\phi} $, which modifies the $tth$ Yukawa coupling. We showed, however, that for a natural underlying NP with $O(1)$ couplings, CPV effects from this top-Higgs operator are expected to be too small - below the expected sensitivity of the LHC and other future high-energy colliders, unless there is yet-undiscovered physics in the sub-TeV region.

We thus conclude that, under the naturality condition of the underlying heavy theory, a measurable CP-violating effect can arise in high-energy scattering processes {\it only} in case (iii) mentioned above, where the NP has sizable, $O(1)$, flavor-changing interactions and the CPV signal is generated purely in the NP sector, i.e., from NP$\times$NP effects and not from SM$\times$NP interference. 

We provide several examples that illustrate our general conclusions, by studying potential CPV from NP physics  in $e^+ e^- \to t \bar t h$ at a future high-energy $e^+e^-$ collider and in top-Higgs associated production at the LHC via $pp \to thW,thj$. 


\bibliographystyle{hunsrt.bst}
\bibliography{mybib2}

\end{document}